\documentclass{IEEEtran}
\usepackage{cite}
\usepackage{amsmath,amssymb,amsfonts}
\usepackage{algorithmic}
\usepackage{graphicx}
\usepackage{textcomp}
\usepackage{algorithmic}
\usepackage{algorithm}
\usepackage{tabularx}
\usepackage{subfigure}

\def\BibTeX{{\rm B\kern-.05em{\sc i\kern-.025em b}\kern-.08em
    T\kern-.1667em\lower.7ex\hbox{E}\kern-.125emX}}
\begin{document}
\title{RIS Assisted Wireless Communication: Advanced Modeling, Simulation, and Analytical Insights}
\author{Xiaocun Zong, Fan Yang,  \IEEEmembership{Fellow, IEEE}, Zhijun Zhang,  \IEEEmembership{Fellow, IEEE}, Shenheng Xu, \IEEEmembership{Member, IEEE} 

and Maokun Li, \IEEEmembership{Fellow, IEEE}

\thanks{This paragraph of the first footnote will contain the date on 
which you submitted your paper for review. It will also contain support 
information, including sponsor and financial support acknowledgment. For 
example, ``This work was supported in part by the U.S. Department of 
Commerce under Grant BS123456.'' }
\thanks{The next few paragraphs should contain 
the authors' current affiliations, including current address and e-mail. For 
example, F. A. Author is with the National Institute of Standards and 
Technology, Boulder, CO 80305 USA (e-mail: author@boulder.nist.gov). }
\thanks{S. B. Author, Jr., was with Rice University, Houston, TX 77005 USA. He is 
now with the Department of Physics, Colorado State University, Fort Collins, 
CO 80523 USA (e-mail: author@lamar.colostate.edu).}
\thanks{T. C. Author is with 
the Electrical Engineering Department, University of Colorado, Boulder, CO 
80309 USA, on leave from the National Research Institute for Metals, 
Tsukuba, Japan (e-mail: author@nrim.go.jp).}}

\maketitle

\begin{abstract}
This article presents a novel perspective to model and simulate reconfigurable intelligent surface (RIS)-assisted communication systems. Traditional methods in antenna design often rely on array method to simulate, whereas communication system modeling tends to idealize antenna behavior. Neither approach sufficiently captures the detailed characteristics of RIS-assisted communication. To address this limitation, we propose a comprehensive simulation framework that jointly models RIS antenna design and the communication process. This framework simulates the entire communication pipeline, encompassing signal generation, modulation, propagation, RIS-based radiation, signal reception, alignment, demodulation, decision, and processing. Using a QPSK-modulated signal for validation, we analyze system performance and investigate the relationship between bit error rate (BER), aperture fill time, array size, and baseband symbol frequency. The results indicate that larger array sizes and higher baseband symbol frequencies exacerbate aperture fill time effects, leading to increased BER. Furthermore, we examine BER variation with respect to signal-to-noise ratio (SNR) and propose an optimal matching-based alignment algorithm, which significantly reduces BER compared to conventional pilot-based alignment methods. This work demonstrates the entire process of RIS communication, and reveals the source of bit errors, which provides valuable insights into the design and performance optimization of RIS-assisted communication systems.  

\end{abstract}


\section{Introduction}
\label{sec:introduction}
\IEEEPARstart{W}{ITH} the widespread deployment and adoption of fifth-generation (5G) wireless communication networks, increasing attention is now being directed toward beyond 5G (B5G) and sixth-generation (6G) technologies. These future networks are anticipated to meet far more stringent requirements, including ultra-high spectral and energy efficiency, microsecond-level latency, and comprehensive full-dimensional network coverage [1], [2]. To address these challenges, enhancements and extensions of existing communication technologies [3], which have already demonstrated their effectiveness, are being proposed. Notable advancements include ultra-massive multiple-input multiple-output (UM-MIMO), ultra-dense networks (UDNs), and terahertz (THz) communication.

However, the deployment of an increasing number of antennas and base stations (BSs), coupled with the use of extremely high carrier frequencies, poses significant challenges, such as higher energy consumption and increased hardware costs. This arises primarily from the requirement for additional power-intensive and costly radio frequency (RF) chains needed for signal conversion. Furthermore, the integration of a large number of active components at such frequencies introduces additional complexities, including inter-user and inter-cell interference, pilot contamination, and pronounced hardware impairments.

To overcome these limitations, cost-effective solutions for wireless communication systems must be explored. Inspired by advancements in metasurfaces and fabrication technologies, reconfigurable intelligent surfaces (RIS) have emerged as a promising solution [4]–[8]. RIS can be described as a programmable array composed of a large number of periodically arranged elements, functioning as a novel type of phased array. Based on feed position, RIS can be categorized into reflective reconfigurable arrays (RRA) and transmissive reconfigurable arrays (TRA). Due to its spatial feeding and integrated phase-tuning techniques, RIS offers a significantly simpler and lower-cost structure compared to conventional phased arrays [9]–[11]. Typically controlled by field-programmable gate arrays (FPGA), RIS enables flexible beamforming, active channel construction, and signal modulation. This flexibility has led to its growing popularity in academia and its adoption in a range of civilian and military communication applications.

From an antenna design perspective, RIS is often modeled using array methods and simulated through electromagnetic simulation software, such as $HFSS$ and $CST$. However, such simulations are limited as they cannot directly depict the characteristics of the received signal when a real communication signal is incident on the RIS. Conversely, in communication system simulations of RIS-assisted networks, the RIS is typically idealized as a simple point or surface. This oversimplification neglects the intricate design details of RIS elements, thereby compromising accuracy and reliability, and failing to capture the real channel characteristics.

Therefore, a critical challenge lies in jointly designing the antenna and communication system. Specifically, it is essential to investigate how communication signals are modulated, how RIS elements manipulate and radiate electromagnetic waves, how multiple signal channels are received and combined, and how signal alignment and demodulation are performed. Addressing these issues in detail will contribute to a more accurate and holistic understanding of RIS-assisted communication systems.

This article proposes a novel method for the joint simulation of antennas and communication systems, integrating the array modeling method with communication signal processing. Using QPSK-modulated signals as an example, the entire process—including communication signal generation, modulation, propagation, RIS control, reception, merging, alignment, dicision, and post-processing—is systematically analyzed. Additionally, the article investigates the relationship between bit error rate (BER) and factors such as aperture fill time, array size, and baseband frequency. To address alignment challenges, an alignment algorithm based on optimal matching is proposed, which achieves a significant reduction in BER compared to traditional pilot-based alignment methods. Finally, the article presents the BER-SNR performance curves of the RIS-assisted communication system, providing valuable insights into system performance and optimization.

\section{Basic Process And Principles}
A communication system usually includes a transmitter, a propagation channel and a receiver. The RIS-assisted communication system introduces RIS to actively build and modulate the wireless channel, as is shown in Fig.1. In general, the transmission equation of RIS-assisted communication system can be modeled as formula (1), the role of RIS is to reconstruct $h$, as shown in Fig.2.
\begin{figure*}[h]
	\centering
	\includegraphics[width=0.9\linewidth]{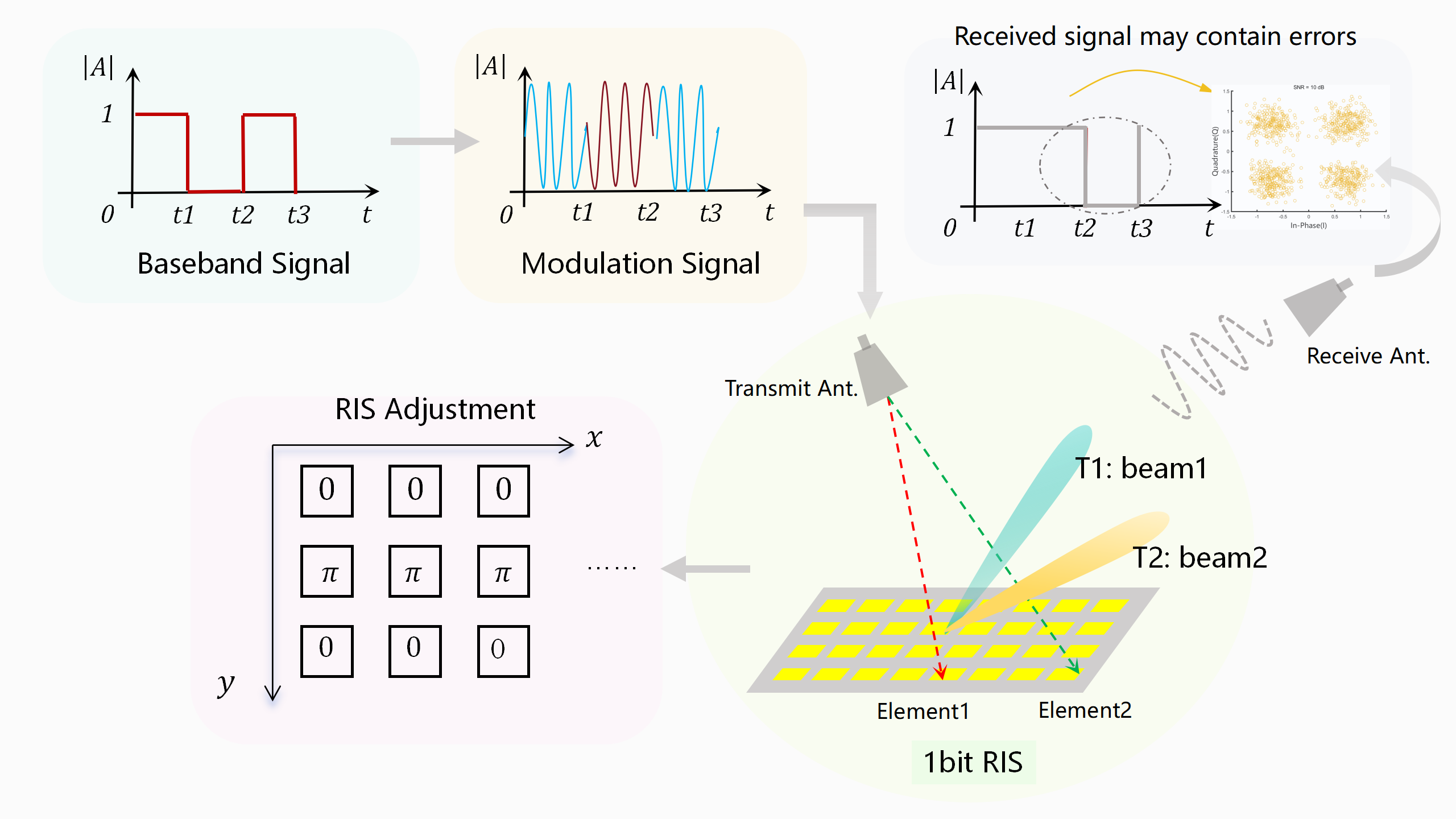}
	\caption{Concept diagram of RIS-assisted wireless communication.}
\end{figure*}
\begin{equation}
	\begin{split}
	r(t) & = \sum_{n=1}^N\sum_{m=1}^M \eta_{mn}s(t-\tau_{mn})\\
        & = \sum_{n=1}^N\sum_{m=1}^M \eta_{mn}e^{j2\pi f_c(t-\tau_{mn})}\\
        & = \sum_{n=1}^N\sum_{m=1}^M \eta_{mn}e^{-j2\pi f_c\tau_{mn}}\cdot e^{j2\pi f_ct}\\
        & = h \cdot s(t)
	\end{split}
\end{equation}

\begin{figure}[h]
	\centering
	\includegraphics[width=0.6\linewidth]{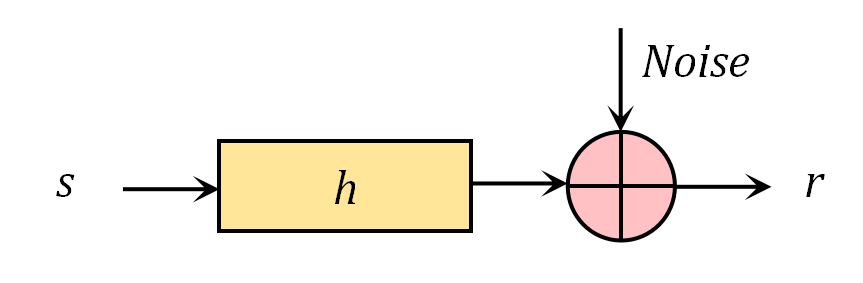}
	\caption{Simplified diagram of channel model.}
\end{figure}

where transmit signal $s(t) = e^{j2\pi f_ct}$, M and N represent the scale of RIS, $\eta_{mn}$ represents the attenuation coefficient of the channel where each element is located, which is $1$ by default in an ideal situation, and $\tau_{mn}$ represents the multipath delay, according to the following analysis, we can know that:

\begin{equation}
\tau_{mn}=\varphi_{mn}^{Q} + \tau_{mn}^{Path}
\end{equation}

\subsection{Baseband Generate and Modulated}
The signal is modulated using QPSK, the input 0/1 bit stream [0,1,1,0] is first converted into bipolar codes [-1,1,1,-1], and then the I/Q two-way signal is separated and the carrier modulation is performed separately. The I/Q signals are each equipped with a sinusoidal modulation signal and modulated to the carrier frequency $f_c$, which is the frequency of the electromagnetic wave used for transmission, as is shown in Fig.3 and Fig.4. 

\begin{figure}[h]
	\centering
	\includegraphics[width=1\linewidth]{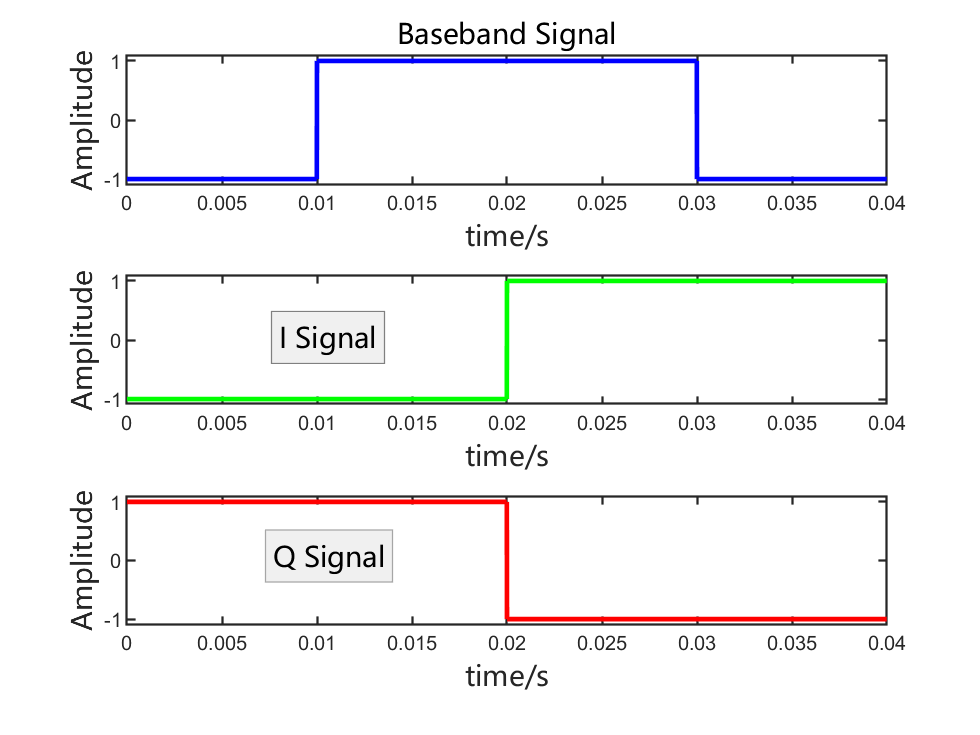}
	\caption{Baseband generate and seperate IQ paths.}
\end{figure}

The generated QPSK signal is fed into the horn and then radiated to each element. The QPSK generation algorithm is shown in Algorithm 1.

\begin{figure}[h]
	\centering
	\includegraphics[width=1\linewidth]{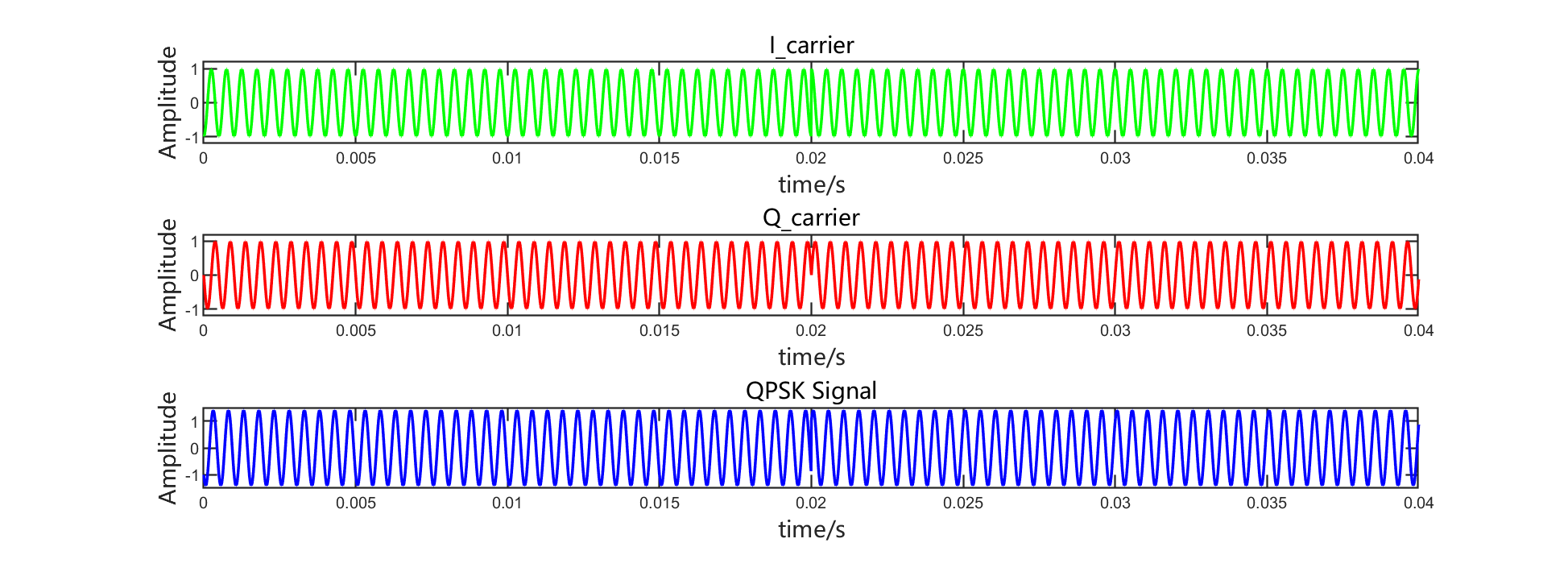}
	\caption{Modulate baseband signal and generate QPSK.}
\end{figure}

\begin{algorithm}[h]
    \caption{QPSK Generation Algorithm}
    \label{alg:AOA}
    \renewcommand{\algorithmicrequire}{\textbf{Input:}}
    \renewcommand{\algorithmicensure}{\textbf{Output:}}
    \begin{algorithmic}[1]
        \REQUIRE 0/1 Random Bitstream  
        \ENSURE QPSK Signal Irradiated to RIS   
        
        \STATE  Generate 0/1 random bitstream,the first two bits are specified as [0,1] and used as pilot sequence: $bitstream = [[0\quad1], randi([0,1], 1, bitnumber-2)]$
        \STATE  0/1 codes are converted to  -1/1 bipolar codes and divided into I/Q paths: $I = bitstream_{2n-1}, Q = bitstream_{2n}$
        \STATE  The I/Q signals are respectively loaded with a carrier wave with the center frequency $f_c$ of RIS
        \STATE  Synthesize QPSK signal: $QPSK = I\times cos(t) + Q\times sin(t)$
        \RETURN QPSK
    \end{algorithmic}
\end{algorithm}

\subsection{RIS Element Adjustment}

The input signal is radiated to each element through the horn, and the phase of each element can be calculated according to the array method. Where $\vec{r}_{fmn}$ is the distance from the feed source to each element, $\vec{r}_{mn}$ is the distance from the irradiation center to each elelement, and $\boldsymbol {u_0}$ is the unit vector of the outgoing beam direction.

\begin{equation}
\varphi_{mn}^{req}=\boldsymbol {k}\cdot (\vec{r}_{fmn}-\boldsymbol {u_0}\cdot\vec{r}_{mn}) + \varphi_{0}
\end{equation}

Quantify the calculated theoretical phase: convert the phase to $(-\frac{\pi}{2}, \frac{3\pi}{2}]$, and the RIS is a 1-bit phase quantization.

\begin{equation}
   \varphi_{mn}^{Q} = \begin{cases}
    0,\quad \varphi_{mn}^{req}\in(0,180^{\circ}] \\
    \pi, \quad \varphi_{mn}^{req}\in(180,360^{\circ}]
    \end{cases}
\end{equation}

A $4\times4$ UPA is used for verification, the signal inputs at 0° normal incidence, outputs at 20° beam, and the focal diameter ratio is 0.9, then the phase of each element can be calculated in Fig.5.
\begin{figure}[H]
	\centering
	\includegraphics[width=0.6\linewidth]{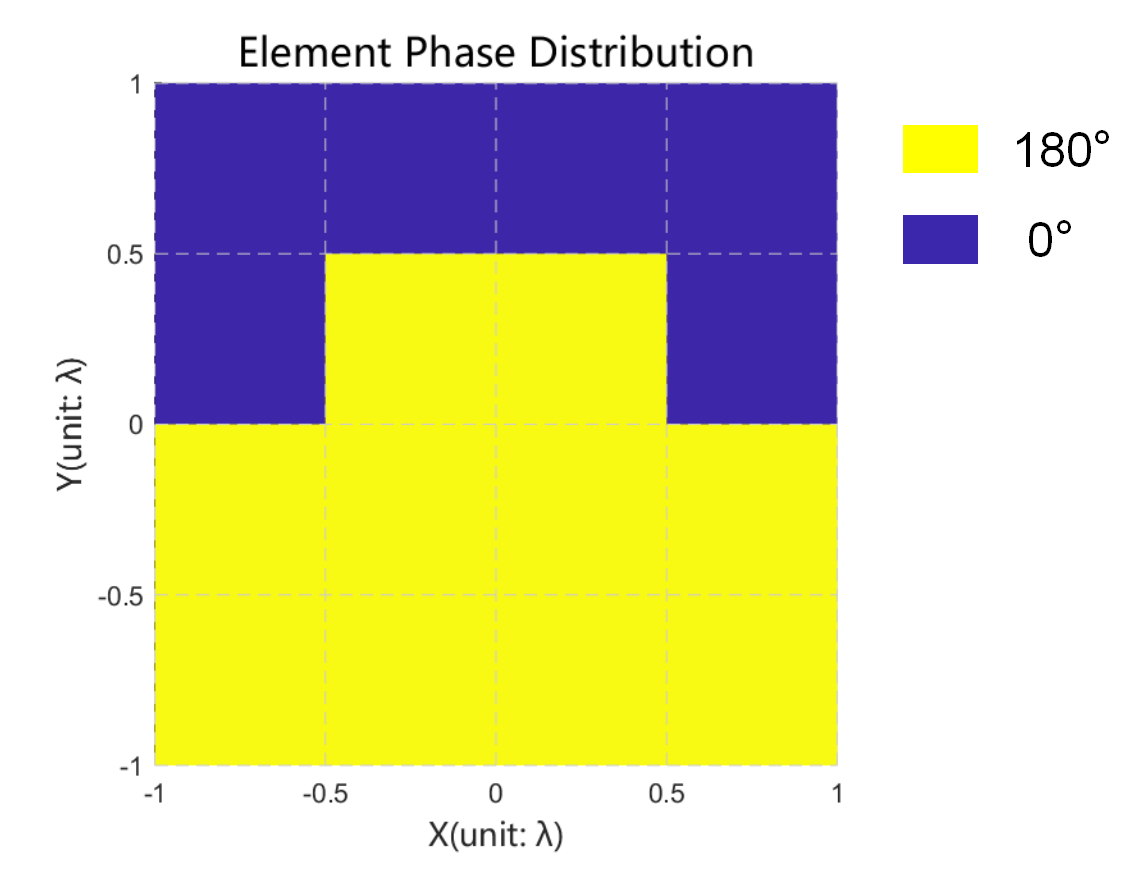}
	\caption{Element Phase Adjustment.}
	\label{fig6}
\end{figure}

Elements at different locations will produce different phase adjustments to the signal. Since it is a 1-bit adjustment, two signals with a phase difference of $180^{\circ}$ after phase adjustment, as is shown in Fig.6. For a signal that adjusts 180° phase, it is equivalent to shifting half a period of the carrier. It should be noted that the 180° phase adjustment cannot directly flip the original waveform, because this is equivalent to flipping the constellation diagram, which will make it impossible to recover the signal. The specific implementation algorithm is shown in Algorithm 2.
\begin{figure}[h]
	\centering
	\includegraphics[width=1\linewidth]{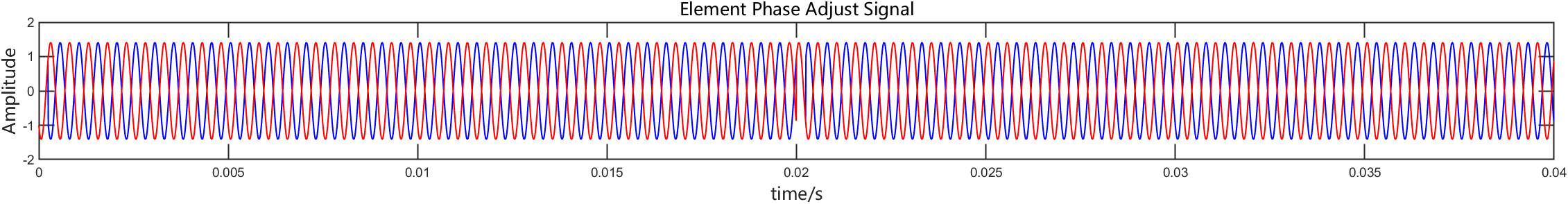}
	\caption{The signal after RIS adjustment is $180^{\circ}$ out of phase.}
\end{figure}

\begin{algorithm}[!h]
    \caption{RIS Adjustment Algorithm}
    \label{alg:AOA}
    \renewcommand{\algorithmicrequire}{\textbf{Input:}}
    \renewcommand{\algorithmicensure}{\textbf{Output:}}
    \begin{algorithmic}[1]
        \REQUIRE Original QPSK Signal  
        \ENSURE Adjusted QPSK Signal   
        
        \FOR{Calculate the quantified phase of each element}
        \IF{ $\varphi_{mn}^{Q} = \pi$}
             \STATE The signal is shifted by half a period of the carrier:
             $ReflectDelay (phase == \pi) = 1/fc/2/dt;$
             $ReflectSignal= circshift(qpsk, ReflectDelay)$;
        \ELSE
             \STATE Signal is not processed;
        \ENDIF
        \ENDFOR
        \RETURN $ReflectSignal$
    \end{algorithmic}
\end{algorithm}

\subsection{Multi-Channel Signal Delay and Merge}

Since the position of each element is different relative to the transmit horn and receive horn, the signal will have different degrees of delay when it incidents the elements and after reflection, which also needs to be taken into account: The specific delay number required is calculated according to formula(5). $rf$ is the distance from the feed source to each element, $reye$ is the distance from the observation point of the received signal to each element, $c$ is the speed of light, and $dt$ is the time of each sampling point. The calculation result is the sampling point that needs to be delayed. The waveform of each signal after element phase adjustment and time delay is shown in Fig.7(a). 

If a constellation diagram is used to represent the signal, as shown in Fig.7(b), it is evident that after applying element phase adjustments and spatial delays, the constellation points corresponding to different paths exhibit varying degrees of rotation.

\begin{equation}
\tau_{mn}^{Path} =  \frac{rf_{mn} + reye_{mn}}{c\times dt}
 \end{equation}
        
\begin{figure}[H]
	\centering
	\includegraphics[width=1\linewidth]{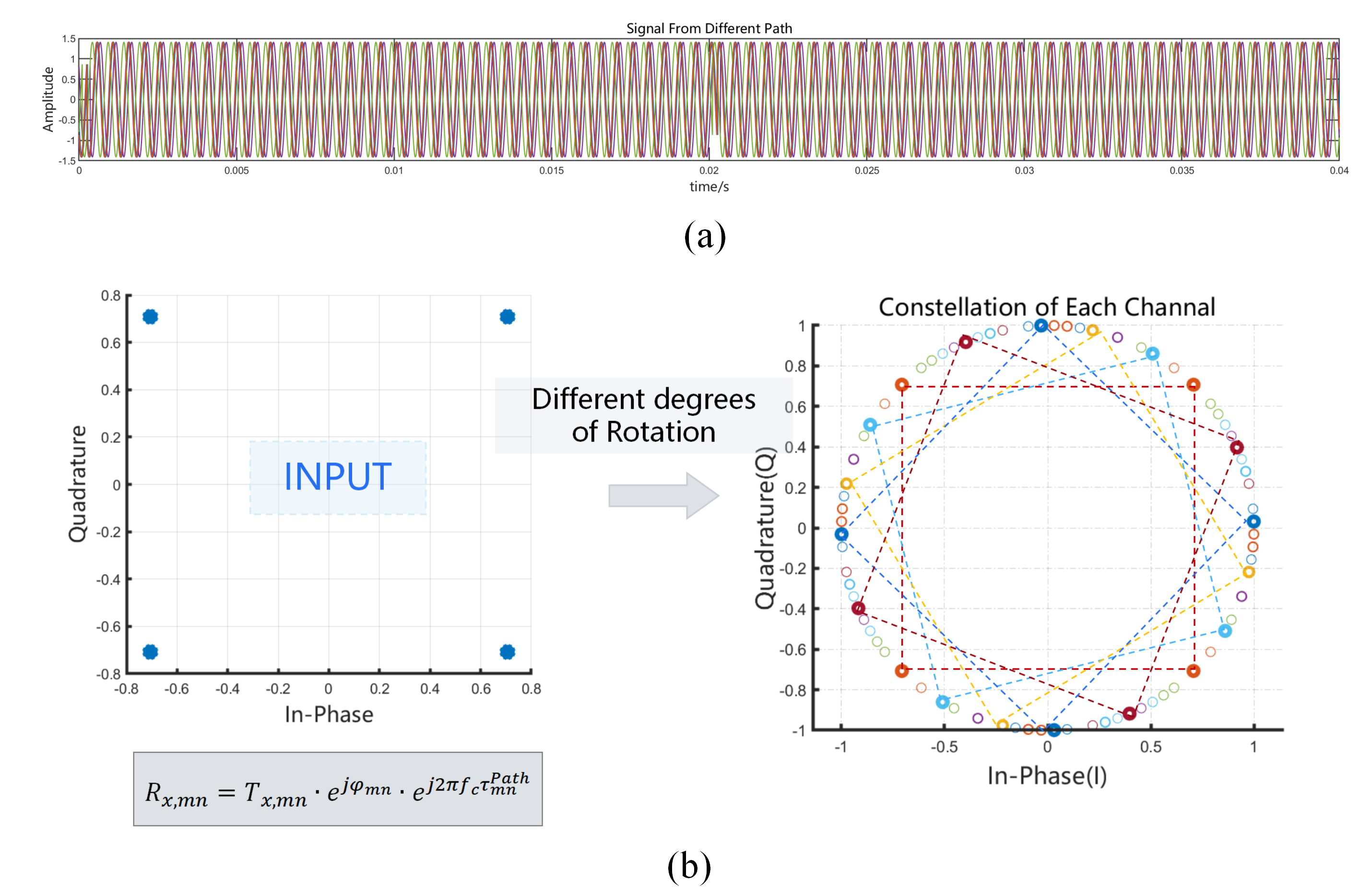}
	\caption{Multi-channel signal delay.}
	\label{fig6}
\end{figure}

During reception, multiple signals must be merged before any subsequent signal processing can occur. Since the signal is real-valued, the signals can be directly superimposed. However, when represented on a constellation diagram, complex numbers need to be added. It is important to note that at the points of code alternation in Fig. 8, the time differences between signals from different paths cause the superposition of different codes, resulting in disturbances and signal distortion at the code transition. This phenomenon is referred to as inter-symbol interference (ISI), which is also the primary contributor to bit error rate (BER).

\begin{figure}[h]
	\centering
	\includegraphics[width=1\linewidth]{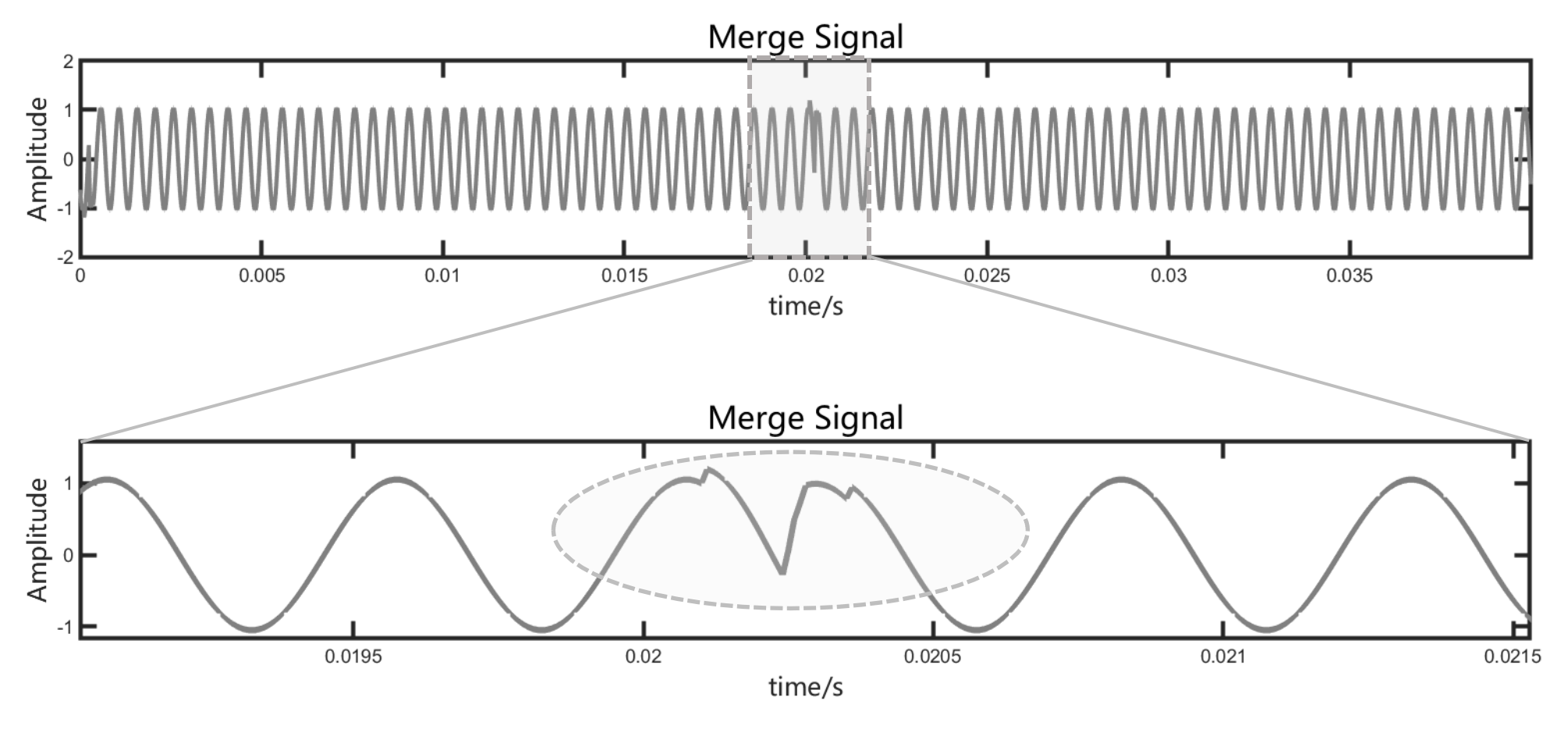}
	\caption{Merge and the signal disturbance at the alternation of codes.}
	\label{fig6}
\end{figure}

\subsection{Alignment, Demodulate and Decision}
Due to the adjustment of the element phase and the delay of multiple signals, the received signal has a certain delay compared with the original signal. This is reflected in the constellation diagram as a rotation of the whole. If alignment is not performed, the demodulated signal must be wrong. Therefore, it is very necessary to align the received signal.

Usually, signal alignment is based on pilot alignment, that is, the sequence with a fixed starting position completes the time alignment, so the entire signal is also synchronized in time. After alignment, the I/Q signals are demodulated and combined separately to recover the original baseband codes.

\begin{figure}[h]
	\centering
	\begin{minipage}{1\linewidth}
		\centering
		\includegraphics[width=0.9\linewidth]{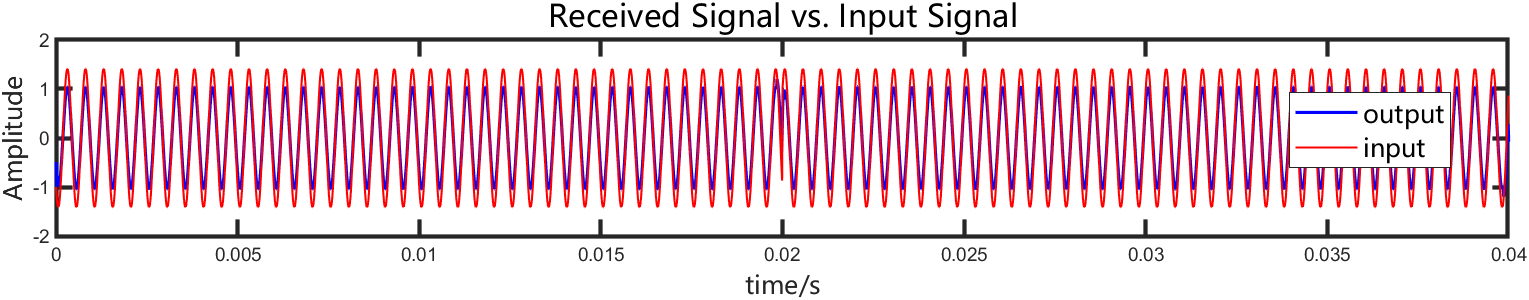}
           \centerline{(a)}
	
		\label{1.(a)}
	\end{minipage}
	
	\begin{minipage}{1\linewidth}
		\centering
		\includegraphics[width=0.9\linewidth]{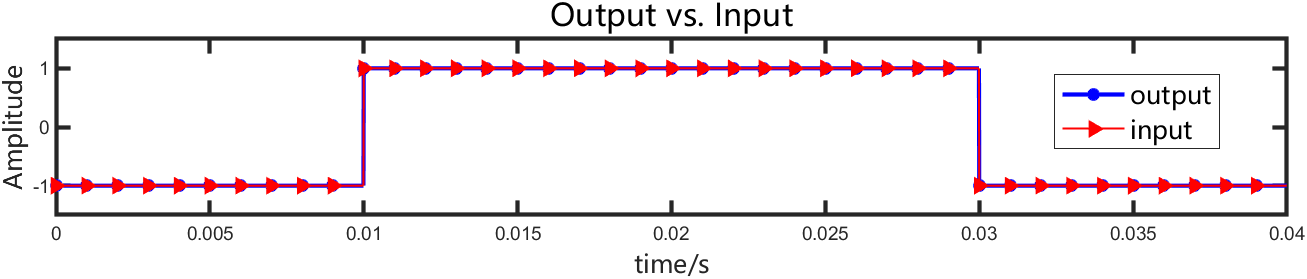}
		\centerline{(b)}
		\label{1.(b)}
	\end{minipage}
 \label{fig.1}
  \caption{ Signal post-processing: (a) Alignment of received signal with original signal, (b) Decision after demodulation.}
\end{figure}

From the recovered baseband signal in Fig.9, we can see that there is no bit error rate compared with the original signal. This is because our array scale is small enough and the symbol frequency is low enough, so there is no aperture fill effect. Subsequent research will be conducted to explore the impact of aperture fill time.

\section{Verification of Algorithm Correctness}

To verify the correctness of the algorithm we proposed, we compare the theoretical bit error rate of QPSK under the AWGN channel model, the equivalent baseband symbol model simulation and our waveform simulation model. We reduced the array size to $1\times1$ to eliminate the multipath effect caused by multiple elements. The theoretical bit error rate of QPSK signal in AWGN channel is shown in the formula(6).

\begin{equation}
P_{b} = \frac{1}{2}erfc(\sqrt{\frac{E_b}{N_0}})
\end{equation}

From the comparison results in Fig.10, we can see that our algorithm and the theoretical value almost perfectly coincide, which verifies the correctness of our algorithm.

\begin{figure}[h]
	\centering
	\includegraphics[width=0.65\linewidth]{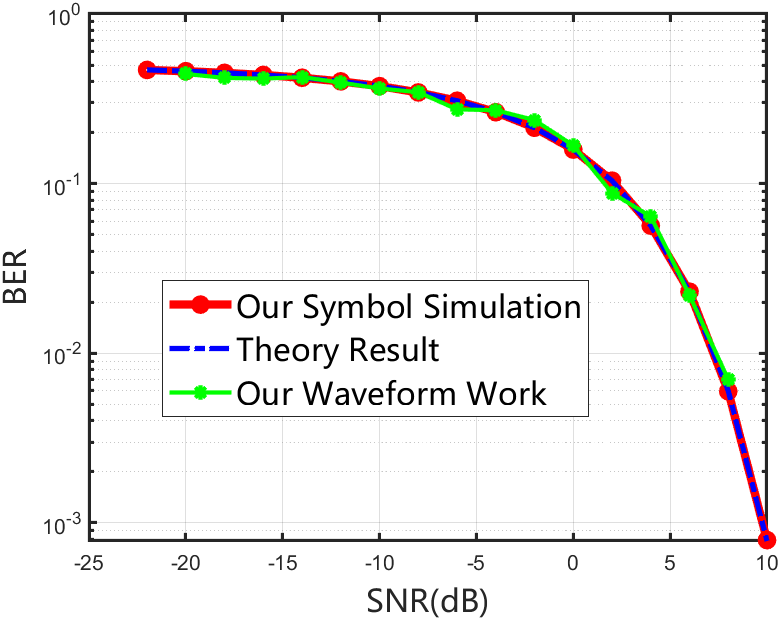}
	\caption{Comparison chart of algorithm correctness verification.}
	\label{fig6}
\end{figure}

\section{Example Verification}
For a real RIS system, the scale of RIS is $16\times16$, the communication signal adopts QPSK modulation, the baseband signal bandwidth is 10MHz, and the carrier frequency is 2GHz. This system is verified on our algorithm to observe the change of its bit error rate characteristics.

The signal is incident along the normal direction of the RIS array, and the outgoing beam direction is $60^{\circ}$. After the communication signal is phase-adjusted and path-delayed by the elements, $16\times16$ channels of signals have different degrees of delay. After calculation, it is determined that the maximum delay for each signal is only 0.03 symbol period, which allows us to neglect the aperture fill effects of each RIS units. As can be seen from the Fig.11, we simulated the constellation diagram under different signal-to-noise ratio conditions, the bit error rate versus signal-to-noise ratio curve of the signal in the AWGN channel is also plotted. There is no bit error rate under the condition of high SNR. We compared waveform-based simulation, symbolic simulation, and a 256×1 MISO simulation to further verify the consistency and correctness of our simulation.

\begin{figure}[h]
	\centering
    \subfigure[]{
	\begin{minipage}{0.75\linewidth}
		\includegraphics[width=0.95\linewidth]{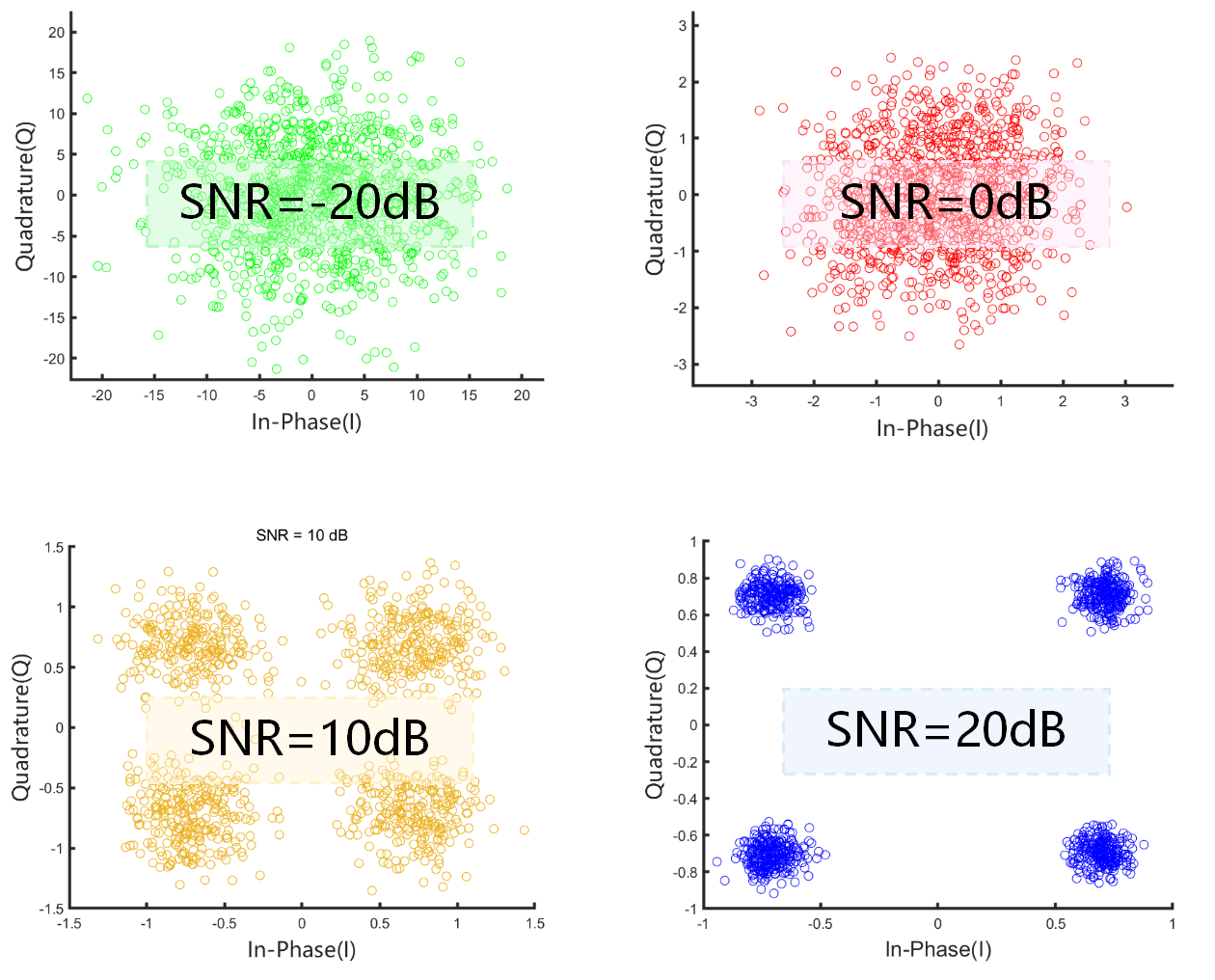}
	\end{minipage}
    }
    \subfigure[]{
	\centering
	\begin{minipage}[(b)]{0.75\linewidth}
		\includegraphics[width=0.95\linewidth]{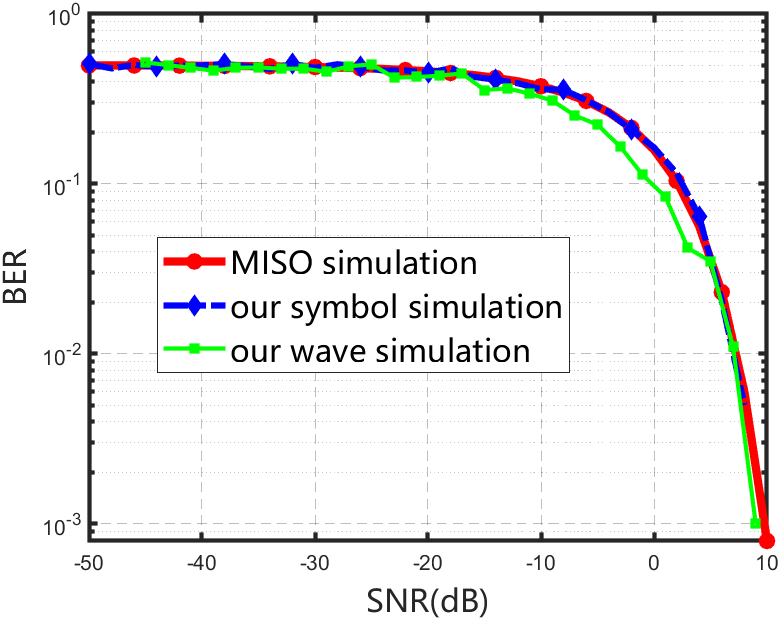}
	\end{minipage}
    }
	\caption{$16\times16$ RIS simulation results:(a) constellation diagram; (b) bit error rate curve.}
	\label{da_chutian}
\end{figure}

\section{Aperture Fill Effect Exploration}
The aperture fill effect refers to the different time delays and phase changes caused by the different phased array elements, which in turn leads to the accuracy of beam scanning , delays and bit errors in signal processing. This has an important impact on radar detection and communication. This chapter will study the effects of array size, baseband frequency $f_{symbol}$, and output beam direction on aperture fill time.

\subsection{Array Size and Baseband Frequency $f_{symbol}$}

When other parameters are kept constant, as the array size increases, the distance between the edge element and the feed source becomes farther, causing the aperture fill effect. The relative distance of the waveform reflected by the edge element becomes longer, resulting in waveform aliasing and bit errors. Keeping other parameters unchanged, when the baseband symbol frequency $f_{symbol}$ increases, the period of each symbol becomes shorter, the edge unit causes aperture crossing, and the relative distance of the reflected waveform becomes larger, resulting in waveform aliasing and bit errors. Essentially, array size and baseband frequency have the same effect on aperture fill effect, both affecting the relative size of symbol delay. The relationship between the two is shown in Fig.12.

\begin{figure*}[h]
	\centering
	\includegraphics[width=0.9\linewidth]{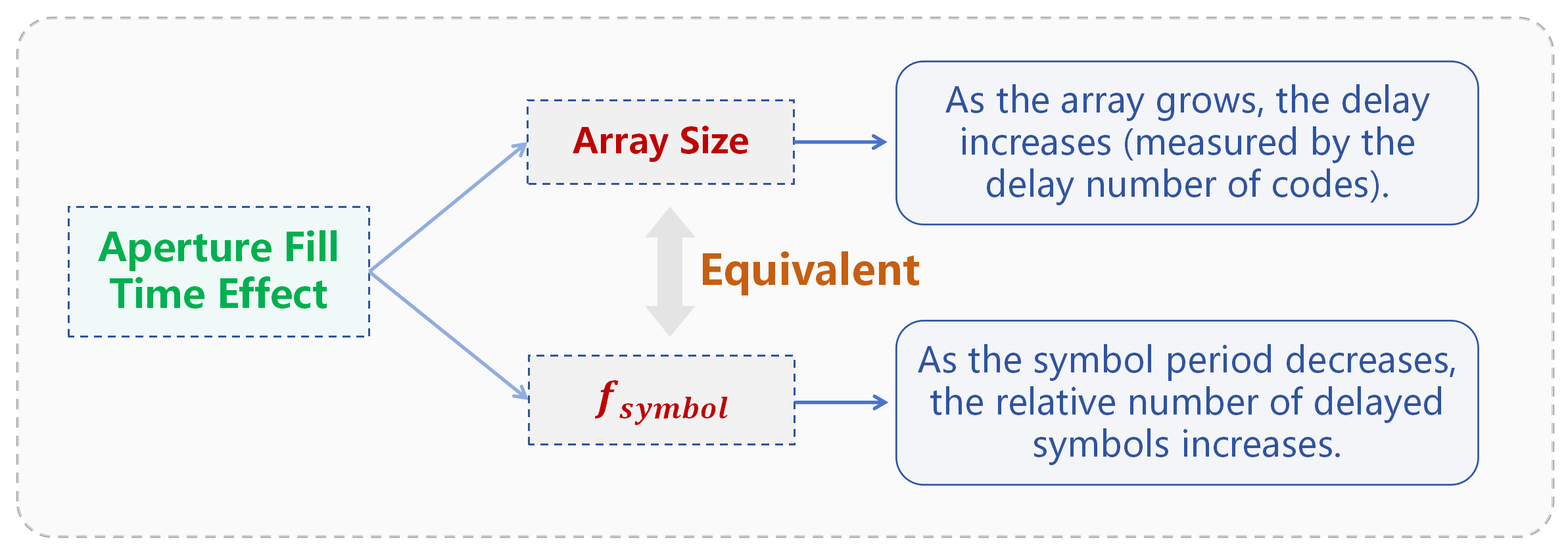}
	\caption{Factors affecting aperture fill time.}
\end{figure*}

In order to speed up the simulation, the array is set to ULA, which is shown in Fig.13, the feed sources are all normal incidence, $0^{\circ}$ beam output.
\begin{figure}[h]
	\centering
	\includegraphics[width=0.7\linewidth]{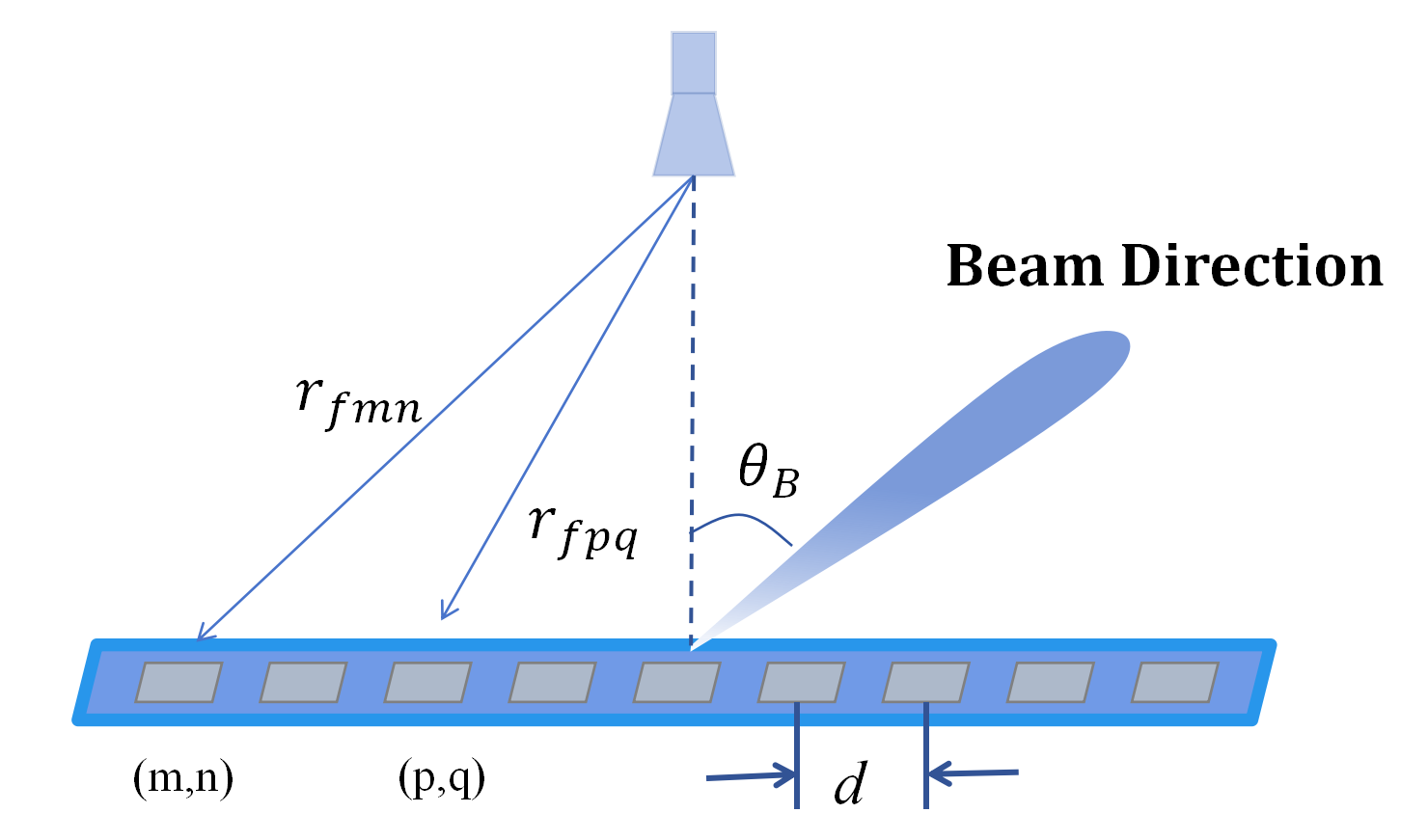}
	\caption{Concept of Uniform Linear Aarry.}
	\label{fig6}
\end{figure}

When the ratio of baseband signal frequency $f_{symbol}$ to carrier frequency $f_{c}$ is 1/2, and the array size is first set to $10\times1$. Since the array is small enough, the total delay of the element at the edge is only 0.3 symbol period, it can be seen in Fig.14(a). After the signal is demodulated using pilot alignment, it can be found that there is \textbf{no BER}.

Keep the signal frequency $f_{symbol}/f_{c}=1/2$ and other parameters unchanged and expand the array size to $100\times1$. It can be seen in Fig.14(b) that the total delay of the unit reflection signal is up to 3 times symbol period, so the interference caused by code element superposition is more serious. A random bit stream with a code length of 200 is used to verify, and \textbf{the BER is 25.5$\textbf{\%}$}. The recovered baseband signal does not overlap with the original signal in many places.

Keep the array size $100\times1$ unchanged, change the signal frequency to $f_{symbol}/f_{c}=1/20$, and keep other parameters unchanged. At this time, because the baseband signal frequency is reduced to 10 times the original, the length of each symbol becomes 10 times the original, so each element's delay (the delayed symbol periods) also becomes 1/10 of the original. Therefore, the superposition of signals will be significantly reduced at this time. In fact, from Fig.14(c) we can see, this situation is completely equivalent to changing the array to $10\times1$ while the frequency remains $f_{symbol}/f_{c}=1/2$. After the signal is demodulated using pilot alignment, it can be found that there is \textbf{no BER}.

\begin{figure*}[h]
	\centering
	\includegraphics[width=1\linewidth]{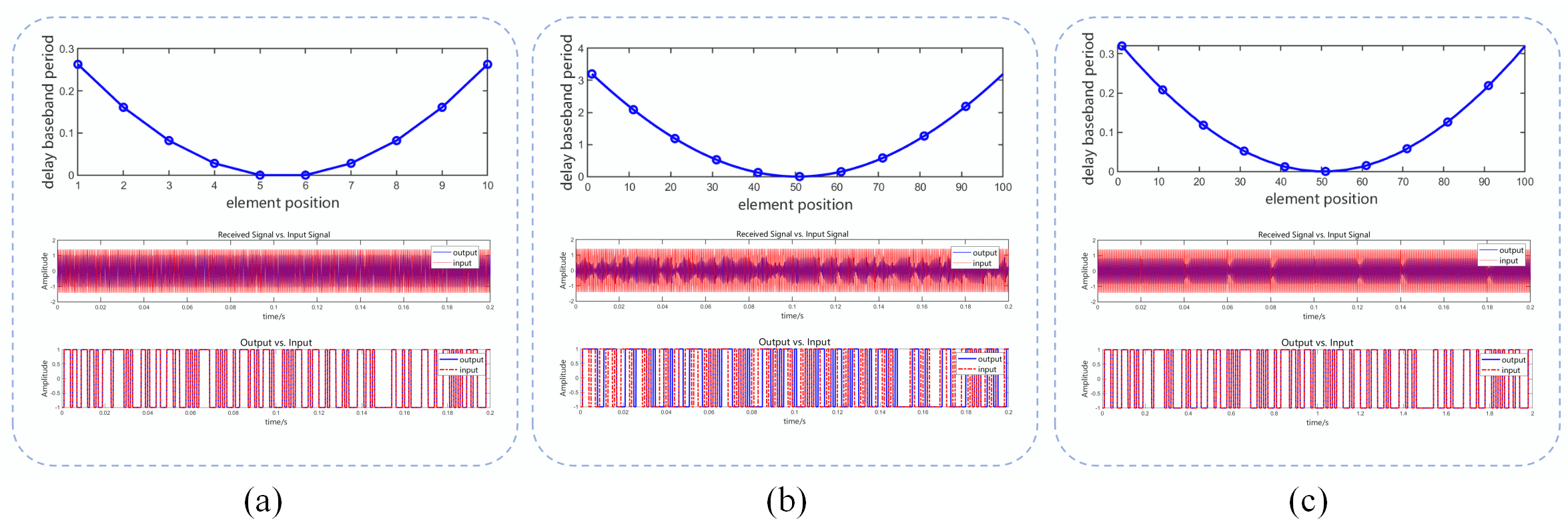}
	\caption{Simulation results for different array sizes and baseband bandwidths at 0°:(a)array size $10\times1$ and $f_{symbol}/f_{c}=1/2$; (b)array size $100\times1$ and $f_{symbol}/f_{c}=1/2$; (c)array size $100\times1$ and $f_{symbol}/f_{c}=1/20$ }
	\label{fig6}
\end{figure*}

\subsection{Effect of Emission Angle}

The emission angle also has a profound impact on aperture fill effect: when the emission is normal, the distribution is asymmetric, the array delay is symmetrically distributed, and the aperture fill phenomenon is not obvious; when the emission is at a large angle, the delay of the left and right elements is asymmetric, with an obvious time difference between the edge element to the equiphase plane. At this time, the impact of aperture fill is very serious, which will further deteriorate the bit error rate.

For the ULA with an array size of $10\times1$ and baseband signal frequency $f_{symbol}/f_{c}=1/2$ , we change the emission angle to $50^{\circ}$. In Fig.15(a), we can find that the total delay of the unit reflected signal is asymmetrically distributed. The maximum total delay is 2 times the baseband symbol period, which is nearly 10 times longer than the $0^{\circ}$ emission. Therefore, the interference caused by code element superposition leads to bit errors. A random bit stream with a code length of 200 is used to verify, and \textbf{the BER is 21$\textbf{\%}$}.

\begin{figure*}[h]
	\centering
	\includegraphics[width=0.8\linewidth]{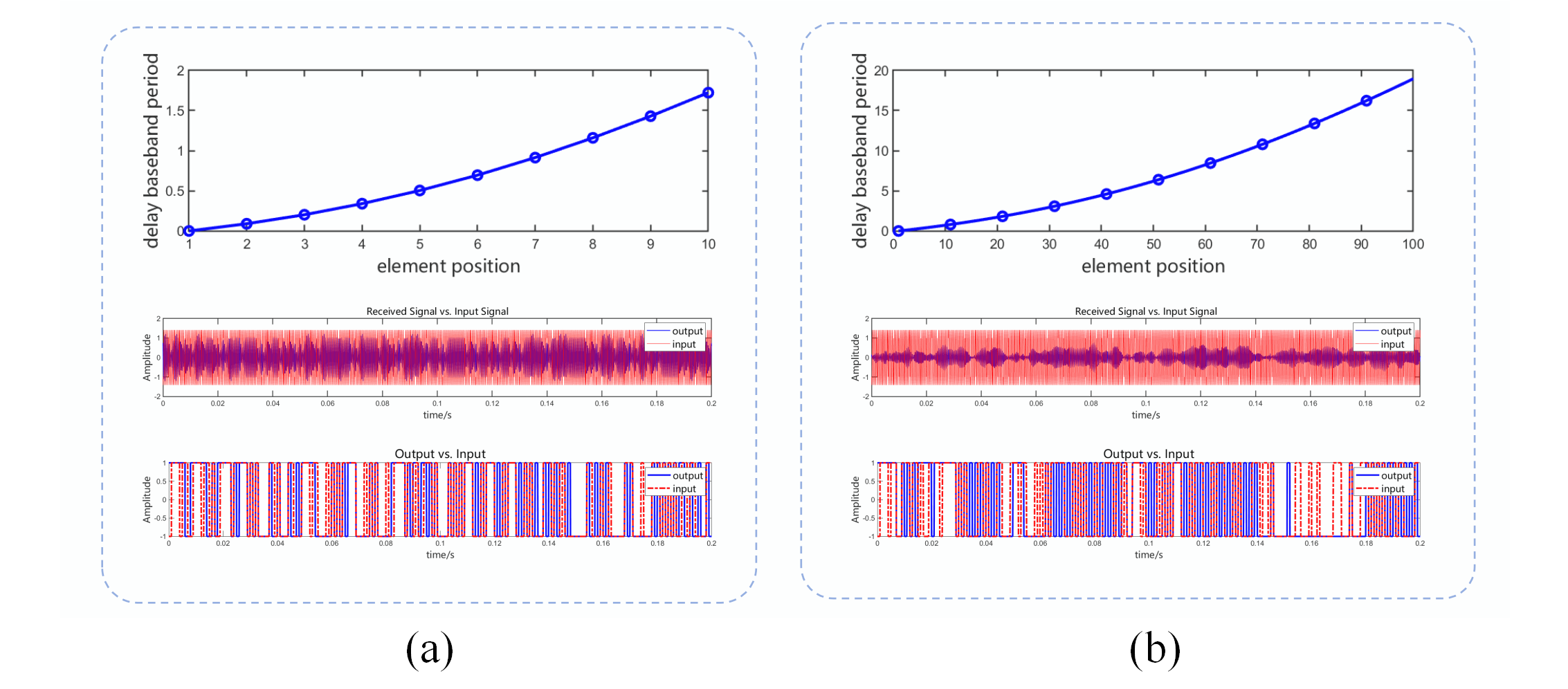}
	\caption{Simulation results for different array sizes and baseband bandwidths at $50^{\circ}$:(a)array size $10\times1$; (b)array size $100\times1$}
	\label{fig6}
\end{figure*}

Keep the signal frequency $f_{symbol}/f_{c}=1/2$ and other parameters unchanged and expand the array size to $100\times1$. It can be seen in Fig.15(b) that the total delay of the unit reflection signal is asymmetrically distributed, and the maximum total delay even reaches 20 times the baseband symbol period. A random bit stream with a code length of 200 is used to verify, and \textbf{the bit error rate increased to 44.5$\textbf{\%}$}.

Verify the situation in a real system: The carrier frequency is set to 2GHz, the baseband signal bandwidth is 100MHz, the communication of $100\times1$ ULA under normal emission and 60° emission conditions respectively, and compare it with $1000\times1$ ULA at normal emission. From the results in Fig.16, we can see that when the normal emission direction is controlled unchanged, when the scale of the ULA increases, the distance from the feed source to the edge element increases, resulting in an increase in the aperture fill time. The aperture fill time of the center element and the edge element are different, resulting in a higher edge delay and bit errors. When the array size is controlled unchanged and the emission angle increases, the distance difference from the edge element to the equiphase plane becomes larger, resulting in an increase in the aperture fill time, causing bit errors, and the bit errors caused by the change in the emission direction are more obvious.
\begin{figure}[H]
	\centering
	\includegraphics[width=0.65\linewidth]{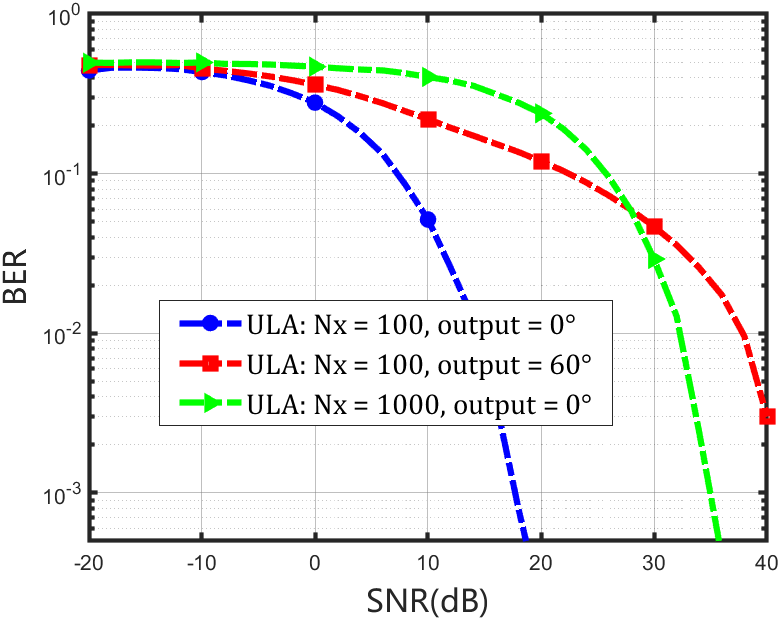}
	\caption{BER under different ULA conditions.}
	\label{fig6}
\end{figure}
\section{Alignment Algorithm Optimization Based on Optimal Matching}

Currently, the mainstream alignment method is based on pilot alignment. For the RIS multipath signal, the pilot of the first arriving signal is used as the reference for alignment. However, this method results in a high bit error rate (BER). To address this limitation, we propose a novel alignment algorithm.
\begin{algorithm}[h]
    \caption{Optimal Matching Alignment Algorithm}
    \label{alg:AOA}
    \renewcommand{\algorithmicrequire}{\textbf{Input:}}
    \renewcommand{\algorithmicensure}{\textbf{Output:}}
    \begin{algorithmic}[1]
        \REQUIRE Merged received signal: R(t)
        \ENSURE Original QPSK Signal: S(t)
        
        \STATE  Calculate the intertwined function $F(\tau)$ of R(t) and S(t): $F(\tau) = R(t) \ast S(t) = \int_{-\infty}^{+\infty}R^{\ast}(t)S(t+\tau)dt $
        \STATE  Find the maximum value of the function $F(\tau)$ and mark its coordinates as $T_{delay} = \tau(max(F(\tau))$
        \STATE  Shift the signal R(t) by a distance $\tau$: $T_{delay} = finddelay(R(t), S(t));R'(t) = circshift(R(t), T_{delay})$;
        \RETURN R'(t)
    \end{algorithmic}
\end{algorithm}

The new alignment algorithm is based on the optimal matching alignment. Calculate the intertwined function of the synthetic signal and the original signal, the maximum value of the interconnected function indicates the time difference between the two signals. The signal after alignment obtained by this alignment method does not correspond to the original signal one by one, but has a certain dislocation. From a certain perspective, this is the average value of the time difference between the pilot of each signal and the original signal pilot.

Since the bit error is essentially caused by the delay caused by multipath, in order to study the change of bit error with delay, the average number of delayed symbol periods in each element channel of the array is used as the independent variable, and the bit error rate is used as the dependent variable. The optimization algorithm is explored to improve the bit error performance.

By plotting the bit error rate versus the average number of delayed symbol periods in Fig.17, we can see that the synchronization algorithm based on optimal matching proposed by us significantly reduces the bit error, it is very effective and robust.
\begin{figure}[h]
	\centering
	\includegraphics[width=0.65\linewidth]{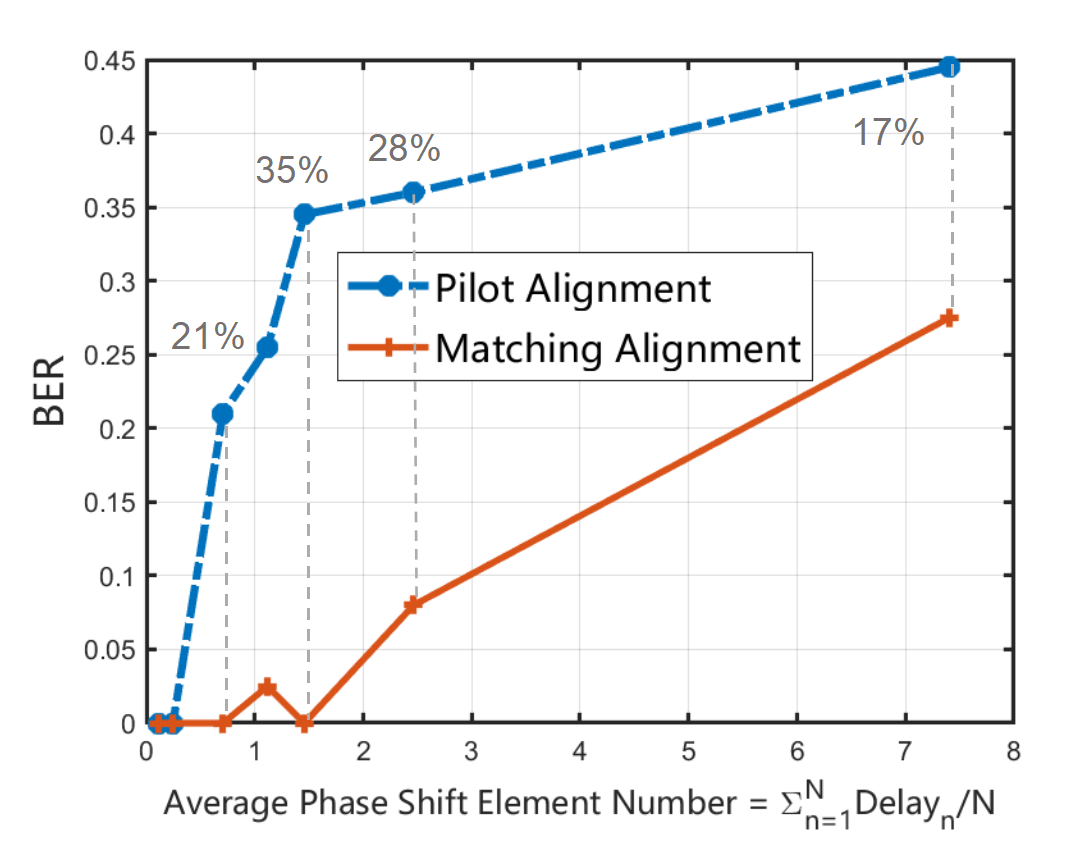}
	\caption{BER versus the average number of shifted symbols.}
	\label{fig6}
\end{figure}

 \section{Conclusions}
This article proposes a novel modeling and simulation method for RIS-assisted communication systems. Using the proposed model, we validate an end-to-end communication system with a $16\times16$ RIS configuration. Based on this method, we investigate the aperture fill time effect of RIS. The results reveal that RIS-assisted wireless communications inherently exhibit multipath issues caused by the aperture fill time of multi-element arrays, which distinguishes them from single-antenna systems. Under ideal conditions, a single antenna does not introduce multipath effects beyond those caused by environmental factors. Furthermore, our study demonstrates that the bit error rate (BER) induced by the aperture fill time effect is influenced by both the array size and the baseband symbol frequency $f_{symbol}$: larger array scales and higher baseband symbol frequencies exacerbate the aperture fill time effect, leading to increased BER. To address this, we propose an alignment algorithm based on optimal matching, which has been experimentally validated to be both effective and robust. This work provides valuable insights and contributes to the optimization of RIS-assisted communication systems.

\end{document}